\documentclass[11pt]{article}

\def\mytitle#1{\setcounter{equation}{0}
\setcounter{footnote}{0}
\begin{flushleft}\Large\textbf{#1}\end{flushleft}
\vspace{0.25cm}}
\def\myname#1{\leftline{{\large #1}}\vspace{-0.13cm}}
\def\myplace#1#2{\small\begin{flushleft}\textit{#1}\\
\texttt{#2}\end{flushleft}}

\usepackage{graphicx}
\begin{document}

\mytitle{Coincidence problem in \textit{f(T)} gravity models}

\vskip0.2cm \myname{Prabir Rudra\footnote{prudra.math@gmail.com}}
\myplace{Department of Mathematics, Asutosh College, Kolkata-700
026, India.}{}

\begin{abstract}
It is well known fact that almost all the recent models of
universe are plagued by the cosmic coincidence problem. In this
assignment we try to probe the role played by torsion in the
current scenario of coincidence and devise a set-up for its
realization. In order to model the scenario, the energy arising
from the torsion component is considered analogous to dark energy.
An interaction between dark energy and dark matter is considered,
which is by far the best possible tool to realize the coincidence.
A set-up is designed and a constraint equation is obtained which
screens the models of \textit{f(T)} gravity that can successfully
accommodate the stationary scenario in its framework, from those
which cannot. Due to the absence of a universally accepted
interaction term introduced by a fundamental theory, the study is
conducted over three different forms of chosen interaction terms.
As an illustration two widely known models of \textit{f(T)}
gravity are taken into consideration and used in the designed
setup. The study reveals that the realization of the coincidence
scenario as well as the role played by torsion in the current
universe is a model dependent phenomenon. It is found that the
first model showed a considerable departure from the stationary
scenario. On the contrary the other four models are perfectly
consistent with our setup and generated a satisfactory stationary
scenario, thus showing their cosmological viability and their
superiority over their counterparts. For the third model
(exponential model) it was seen that the cosmological coincidence
is realized only in the phantom regime. For the fourth
(logarithmic model) and the fifth models, we see that the the
stationary scenario is attained for negative interaction values.
This shows that the direction of flow must be from dark energy to
dark matter unlike the previous models. Under such circumstances
the universe will return from the present energy dominated phase
to a matter dominated phase.

\end{abstract}

\section{Introduction}
At the turn of the last century the incompatibility of General
Relativity (GR) came into light when cosmological observations
from Ia supernovae, CMBR via WMAP, galaxy redshift surveys via
SDSS indicated that the universe is going through an accelerated
expansion of late (Perlmutter, S. et al. 1999; Spergel, D. N. et
al. 2003; Bennett, C. L. et al. 2003; Tegmark, M. et al. 2004;
Allen, S. W. et al. 2004). Since no possible explanation of this
phenomenon could be given in the framework of Einstein's GR, a
proper modification of the theory was required that will
successfully incorporate the late cosmic acceleration. Two
different approaches regarding this are widely found in
literature.

Cosmic acceleration can be phenomenally attributed to the presence
of a mysterious negative energy component popularly known as
\textit{dark energy (DE)} (Riess, A. G. et al. 2004). In this case
the modification is brought about on the right hand side of the
Einstein's equation, i.e. in the matter sector of the universe.
The contribution of DE to the energy sector of the universe is
$\Omega_{d}=0.7$. In due course various candidates for DE began to
appear in the scene. Some of the popular ones worth mentioning are
Chaplygin gas models (Kamenshchik, A. et. al. 2001; Gorini, V. et.
al. 2003), Quintessence Scalar field (Ratra, B. et. al. 1988),
Phantom energy field (Caldwell, R. R. 2002), etc. All these models
violate the strong energy condition i.e., $\rho+3p<0$, thus
producing the observed cosmic acceleration.

The alternative approach is based on the modification of the
gravity sector of GR, thus giving birth to modified gravity
theories. A universe associated with a tiny cosmological constant,
i.e. the $\Lambda$CDM model served as a prototype for this
concept. Although the model could satisfactorily explain the
recent cosmic acceleration and passed a few solar system tests as
well, yet, detailed diagnosis revealed that the model was
paralyzed with a few cosmological problems. The two major problems
that have crippled the model over the last decade are the Fine
tuning problem (FTP) and the Cosmic Coincidence problem (CCP). The
former refers to the large discrepancy between the observed values
and the theoretically predicted values. There have been many
attempts to solve this problem. The most impressive attempt was
probably undertaken by Weinberg in (Weinberg, S. 1989). The
solutions are basically based on the fact that the cosmological
constant may not assume an extremely small static value at all
times during the evolution of the universe, but its nature should
be rather dynamical (Bisabr, Y. 2010). As a result alternative
modifications of gravity was sought for. Some of the popular
models of modified gravity that came into existence in recent
times are loop quantum gravity (Rovelli, C. 1998; Ashtekar, A. et.
al. 2004), Brane gravity (Brax, P. et. al 2004; Maartens, R. 2000,
Maartens, R. 2004), \textit{f(R)} gravity (Kerner, R. 1982;
Allemandi, G. 2004; Carroll, S. M. et. al. 2005), \textit{f(T)}
gravity (Linder, E. V. 2010; Li, M. et. al. 2011; Miao, R. et. al.
2011; Li, B. et. al. 2011), etc. In recent times a lot of work has
been done in modified gravity theories, specially $f(R)$, $f(T)$,
$f(R,T)$, etc. models (Zubair, M. et. al. 2015a; Zubair, M. 2015b,
Waheed, S. et. al. 2015; Harko, T. et. al. 2011; Sarkar, S. et.
al. 2013; Sharif, M. et. al. 2013a, 2013b; Noureen, I. et. al.
2015).

In this work we will consider \textit{f(T)} model as the theory of
gravity. $f(T)$ gravity is an alternative theory for GR, defined
on the Weitzenbock non-Riemannian manifold. In this framework
curvature is replaced by torsion. The formation is basically based
on the division of the manifold into two separate but connected
parts, one of which has a Riemannian structure with a definite
metric and the other one has a non-Riemannian structure with
torsion. The non-Riemannian part is based on a tetrad basis
defining a Weitzenbock spacetime. The basis of this model was
first laid by Einstein in his Teleparallel equivalent of general
relativity (TEGR). At that time the purpose of the model was to
unify electromagnetism and gravity. If we consider $f(T) = T$, the
theory reduces to teleparallel gravity (Hayashi, K. et. al. 1979;
Hehl, F. et. al. 1976). In (Hayashi, K. et. al. 1979), it has been
shown that with a linear form \textit{f(T)}, it is possible to
satisfy the standard solar system tests, thus establishing the
viability of the model. In spite of the success of the linear
\textit{f(T)} model, unfortunately further development took a long
time coming. It was not before 2007 that Ferraro et al (Ferraro,
R. et. al. 2007, 2008) introduced a general model of $f(T)$
gravity. From then, there have been extensive work to enrich the
theory. Here it is worth mentioning that Birkhoff's theorem was
studied in this gravity by Meng et al (Meng, X. et. al. 2011). The
authors in (Zheng, R. et. al. 2011) investigated perturbation in
$f(T)$ and found that the growth of perturbations in $f(T)$
gravity is much slower than that in GR. Bamba et al (Bamba, K. et.
al. 2011) studied the evolution of equation of state parameter and
phantom crossing in $f(T)$ model.

In (Bisabr, Y. 2010) Bisabr studied cosmological coincidence
problem in the background of \textit{f(R)} gravity. Motivated by
Bisabr's work, we dedicate the present assignment to the study the
effect of torsion arising from \textit{f(T)} gravity models in the
present coincidence scenario and devise a set-up for its
realization. The paper is organized as follows: In section 2, the
basic equations of \textit{f(T)} gravity is presented. We address
the coincidence problem in section 3 and the paper ends with some
concluding remarks in section 4.

\section{Basic Equations of \textit{f(T)} Gravity}

A suitable form of  action for $f(T)$ gravity in Weitzenbock
spacetime is  (Ferraro, R. et. al. 2007, 2008)
\begin{equation}\label{1}
S=\frac{1}{2\kappa^2}\int d^{4}x~ e~\left[f(T)+L_{m}\right]
\end{equation}

Here $e=det(e^{i}_{\mu})=\sqrt{-g}$, $\kappa^2=8\pi G$ and
$e^{i}_{\mu}$ is the tetrad (vierbein) basis. The dynamical
quantity of the model is the vierbein $e^{i}_{\mu}$ and $L_m$ is
the matter Lagrangian. Taking the variation of the action
(\ref{1}), with respect to the vierbein $e^{i}_{\mu}$, the
modified Friedmann equations in the spatially flat FRW universe
can be obtained as,

\begin{equation}\label{2}
H^2=\frac{\kappa^2}{3}\left(\rho_m+\rho_T\right)
\end{equation}
\begin{equation}\label{3}
2\dot{H}+3H^{2}=-\kappa^{2}\left(p_{m}+p_{T}\right)
\end{equation}
where
\begin{equation}\label{4}
\rho_T=\frac{1}{2\kappa^2}\left(2Tf_{T}-f-T\right)
\end{equation}
\begin{equation}\label{5}
p_T=-\frac{1}{2\kappa^2}\left[-8T\dot {H}f_{TT}+\left(2T-4\dot
H\right)f_T-f+4\dot{H}-T\right]
\end{equation}
and
\begin{equation}\label{6}
T=-6H^{2}
\end{equation}
Here the subscript $T$ indicates derivative with respect to the
torsion scalar $T$ and obviously $H$ denotes the Hubble parameter.
$\rho_{m}$ and $p_{m}$ represents the energy density and pressure
of the matter content of the universe whereas $\rho_{T}$ and
$p_{T}$ represents the density and pressure contributions of the
scalar torsion.

The energy conservation equations are given by,
\begin{equation}\label{7}
\dot{\rho_{m}}+3H\rho_{m}=Q
\end{equation}
\begin{equation}\label{8}
\dot{\rho_{T}}+3H\left(1+\omega_{T}\right)\rho_{T}=-Q
\end{equation}

Here $\omega_{T}=\frac{p_{T}}{\rho_{T}}$ is the EoS parameter of
the torsion scalar and $Q$ is the interaction between the matter
and the torsion sector of the universe.

The torsion EoS parameter is defined as (Wu, P. et. al. 2010,
Karami, K. et. al. 2012)
\begin{equation}\label{10}
\omega_{T}=\frac{p_{T}}{\rho_{T}}=-1-\frac{\dot{T}}{3H}\left(\frac{2Tf_{TT}+f_{T}-1}{2Tf_{T}-f-T}\right)
\end{equation}
For a de-Sitter universe (empty), $\dot{H}=\dot{T}=0$. So
eqn.(\ref{10}) gives $\omega_{T}=-1$, which corresponds to the
$\Lambda CDM$ model.

Using eqns.(\ref{2}), (\ref{4}) and (\ref{6}), we calculate the
matter density as,
\begin{equation}\label{11}
\rho_{m}=\frac{1}{2\kappa^{2}}\left(f-2Tf_{T}\right)
\end{equation}
Here we will consider pressureless matter. So using $p_{m}=0$ in
the eqn.(\ref{3}) and then using eqns.(\ref{2}) (\ref{5}) and
(\ref{10}) we obtain,
\begin{equation}\label{12}
\dot{T}=3H\left(\frac{f-2Tf_{T}}{2Tf_{TT}+f_{T}}\right)
\end{equation}
Using the above expression of $\dot{T}$ in eqn. (\ref{10}), we
arrive at the final expression for EoS parameter for torsion
scalar as,
\begin{equation}\label{13}
\omega_{T}=-\frac{f/T-f_{T}+2Tf_{TT}}{\left(f_{T}+2Tf_{TT}\right)\left(f/T-2f_{T}+1\right)}
\end{equation}
The deceleration parameter is given by
\begin{equation}\label{14}
q=-1-\frac{\dot{H}}{H^{2}}
\end{equation}
Using eqns.(\ref{6}) and (\ref{12}), we get the expression for
deceleration parameter in \textit{f(T)} gravity as,
\begin{equation}\label{15}
q=2\left(\frac{f_{T}-Tf_{TT}-3f/4T}{f_{T}+2Tf_{TT}}\right)
\end{equation}
For teleparallel equivalent of general relativity (TEGR),
$\textit{f(T)}=T$. Using this in the above equation we get
$q=0.5$, which corresponds to decelerating universe, which in turn
points towards a matter dominated scenario.

\section{The Cosmic Coincidence Problem}
The cosmic coincidence problem has been a serious issue in recent
times regarding various dark energy models. Recent cosmological
observations have shown that densities of the matter sector and
the DE sector of the universe are almost identical in the late
universe. We know that the matter and the energy component of the
universe have evolved independently from different mass scales in
the early universe, then how come they reconcile to the same mass
scales in the late universe! This is our problem. Almost all the
DE models known till date suffer from this phenomenon.

There have been numerous attempts to solve the coincidence
problem. Among them the most impressive one is the concept of a
suitable interaction between the matter and the energy components
of the universe, as used in the conservation equations (\ref{7})
and (\ref{8}). Here we consider that the two sectors of the
universe have not evolved independently from different mass
scales. But they interact with each other, thus allowing a mutual
flow between the two components. As a result, the densities of the
two components coincide in the present universe. Although the
concept seems to be promising yet there remains a problem. There
is no universally accepted interaction term, introduced by a
fundamental theory.

\subsection{Choice of Interaction term}
Due to the unknown nature of both dark energy and dark matter, it
is not possible to derive an expression for interaction ($Q$)
using the first principles. So in such a situation, one is
expected to use logical reasoning and propose various feasible
expressions for $Q$. Observing the domination of dark energy in
late times the best possible argument is to consider $Q$ to be
small and positive. A large negative $Q$ will see the universe
dominated by dark energy from the early times, thus leaving no
scope for the condensation of galaxies. The most obvious choice
for interaction should be the energy densities multiplied by the
hubble parameter, because it is both physically and dimensionally
justified. So $Q=Q(H\rho_{m}, H\rho_{de})$, where $\rho_{de}$ is
the dark energy density. Since here we are not adding any dark
energy by hand, so the torsion component of energy, $\rho_{T}$
will replace $\rho_{de}$. This leads us to three basic forms of
interactions as given below (del Campo, S. et. al. 2009):
\begin{equation}\label{int}
\Gamma-model:~ Q=3\Gamma
H\left(\rho_{m}+\rho_{T}\right),~~~~~~~~\zeta-model:~ Q=3\zeta
H\rho_{m},~~~~~~~~\eta-model:~ Q=3\eta H\rho_{T}
\end{equation}
where $\Gamma$, $\zeta$ and $\eta$ are the coupling parameters of
the respective interaction models.

It is worth mentioning that due to its simplicity the most widely
used interaction model is the $\zeta$-model and is available
widely in literature (Berger, M. S. et. al. 2006, del Campo, S.
et. al. 2009, Rudra, P. et. al. 2012, Jamil1).

\subsection{Coincidence in presence of torsion: The set-up}
In this note we try to probe the role played by the torsion
component of the gravity theory in producing the present
coincidence scenario. \textit{f(T)} gravity has evolved over the
past few years as a major candidate of modified gravity theory
satisfying all the solar system tests. \textit{f(T)} gravity is
itself self competent in producing the late cosmic acceleration
without resorting to any forms of dark energy. Therefore we do not
consider any separate dark energy components in the present study,
but the equivalent energy evolving from the torsion component of
the \textit{f(T)} gravity is considered as the dark energy. We
consider the ratio of the densities of matter and dark energy as,
$r\equiv \rho_{m}/\rho_{T}$. Our aim is to search for some
appropriate forms of the function \textit{f(T)} that produces a
stationary value of the ratio of the component densities, $r$. The
time evolution of $r$ is as follows,
\begin{equation}\label{16}
\dot{r}=\frac{\dot{\rho_{m}}}{\rho_{T}}-r\frac{\dot{\rho_{T}}}{\rho_{T}}
\end{equation}

Using eqns. (\ref{7}), (\ref{8}) and (\ref{16}), we obtain
\begin{equation}\label{17}
\dot{r}=3Hr\omega_{T}+\frac{Q}{\rho_{T}}\left(1+r\right)
\end{equation}
Using eqns. (\ref{4}), (\ref{13}) and (\ref{15}) in
eqn.(\ref{17}), we get,
\begin{equation}\label{18}
\dot{r}=3Hr\left[\frac{3r}{2\left(r+1\right)
\left(-12H^{2}f_{TT}+f_{T}\right)}-\left(\frac{f+12H^{2}f_{T}}
{-12H^{2}f_{TT}+f_{T}}\right)\left(\frac{-12H^{2}f_{TT}+f_{T}-1}
{-12H^{2}f_{T}-f+12H^{2}}\right)+q\right]+\frac{2\kappa^{2}\left(1+r\right)Q}{-12H^{2}f_{T}-f+6H^{2}}
\end{equation}
Now in order to comply with observations, it is required that
universe should approach a stationary stage, where either $r$
becomes a constant or evolves slower than the scale factor. In
order to satisfy this $\dot{r}=0$ in the present epoch. It leads
to the following equation,
\begin{equation}\label{19}
g(f,H,r_{s},q)=0
\end{equation}
where

$$g(f,H,r_{s},q)\equiv
3Hr_{s}\left[\frac{3r_{s}}{2\left(r_{s}+1\right)
\left(-12H^{2}f_{TT}+f_{T}\right)}-\left(\frac{f+12H^{2}f_{T}}
{-12H^{2}f_{TT}+f_{T}}\right)\left(\frac{-12H^{2}f_{TT}+f_{T}-1}
{-12H^{2}f_{T}-f+12H^{2}}\right)+q\right]$$
\begin{equation}\label{20}
+\frac{2\kappa^{2}\left(1+r_{s}\right)Q}{-12H^{2}f_{T}-f+6H^{2}}
\end{equation}
where $r_{s}$ is the value of $r$ when it takes a stationary
value.

Now eqn.(\ref{19}) gives us a constraint which can be used to find
suitable \textit{f(T)} functions consistent with a late time
stationary scenario of energy densities. It can also be used to
check whether a particular \textit{f(T)} model fits the stationary
scenario or not. For a given redshift $z_{0}$ at a sufficiently
late time, the corresponding contemporary parameters will be given
by $r_{s}(z_{0})$, $H(z_{0})$ and $q(z_{0})$. At sufficiently late
times eqn.(\ref{19}) can be rewritten as,
\begin{equation}\label{21}
g(f_{0},H_{0},r_{s0},q_{0})=0
\end{equation}
where

$$g(f_{0},H_{0},r_{s0},q_{0})\equiv
3H_{0}r_{s0}\left[\frac{3r_{s0}}{2\left(r_{s0}+1\right)
\left(-12H_{0}^{2}f_{0}''+f_{0}'\right)}-\left(\frac{f_{0}+12H_{0}^{2}f_{0}'}
{-12H_{0}^{2}f_{0}''+f_{0}'}\right)\left(\frac{-12H_{0}^{2}f_{0}''+f_{0}'-1}
{-12H_{0}^{2}f_{0}'-f_{0}+12H_{0}^{2}}\right)+q\right]$$
\begin{equation}\label{22}
+\frac{2\kappa^{2}\left(1+r_{s0}\right)Q}{-12H_{0}^{2}f_{0}'-f_{0}+6H_{0}^{2}}
\end{equation}
where $r_{s0}=r_{s}(z_{0})$ and $f_{0}$, $f_{0}'$, $f_{0}''$
represents the late-time configurations of $f$, $f_{T}$ and
$f_{TT}$.

As far as $q$ is concerned, we start from the best fit
parametrization obtained directly from observational data. Here we
use a two parameter reconstruction function for $q(z)$ (Gong, Y.
G. et. al. 2006, 2007)
\begin{equation}\label{23}
q(z)=\frac{1}{2}+\frac{q_{1}z+q_{2}}{\left(1+z\right)^{2}}
\end{equation}
On fitting this model to Gold data set, we get
$q_{1}=1.47_{-1.82}^{+1.89}$ and $q_{2}=-1.46\pm 0.43$ (Gong, Y.
G. et. al. 2007). We consider $z_{0}=0.25$ and using these values
in eqn.(\ref{23}), we get $q_{0}\approx -0.2$. From recent
observations, we obtain $r_{0}\equiv
\frac{\rho_{m}(z_{0})}{\rho_{T}(z_{0})}\approx \frac{3}{7}$
(Zlatev, I. et. al. 2000; Wei, H. et. al. 2005; Yang, G. et. al.
2005).

Now in order to illustrate the present scheme of work we will
consider some specific models and use them in the constraint
eqn.(\ref{21}). The models that will satisfy the constraint
equation will be cosmologically more acceptable models since they
admit the cosmic coincidence in their framework. In the light of
this, we may possibly be able to rule out some of the known models
which will not satisfy the constraint, thus becoming inconsistent
with observations.

\subsection{Illustration}

\subsubsection{Model 1}

This model was proposed by Abdalla et al in 2005 (Abdalla, M. C.
B. et. al. 2005). It is given by,
\begin{equation}\label{24}
f(T)=\beta T+\gamma T^{n}
\end{equation}
where $\beta$, $\gamma$ and $n>0$ is a constants.

We fit the above model in the constraint eqn.(\ref{21}) and plot
the function $g(f_{0},H_{0},r_{s0},q_{0})$ against the parameter
$n$ in figs. (1a), (1b) and (1c) for $\Gamma$, $\zeta$ and $\eta$
interaction terms respectively. From the figures it is seen that
the function $g$ never attains the zero level. So the constraint
(\ref{21}) is not satisfied for any values of $n$. Hence the model
does not admit a late time stationary ratio of energy densities.
Therefore it is not a cosmologically viable model as far as the
latest observational data is concerned.

\vspace{2mm}
\begin{figure}
~~~~~~~~~\includegraphics[height=2in]{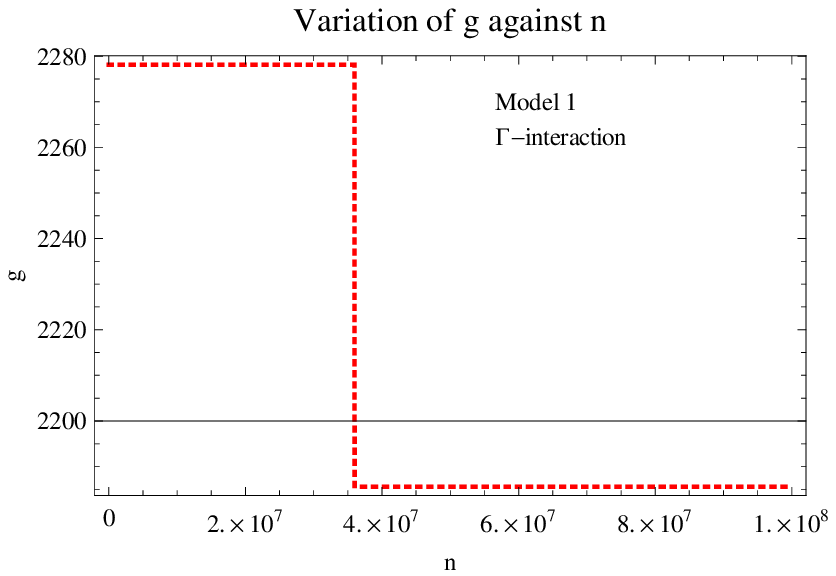}~~~~~~~~~~~~~~\includegraphics[height=2in]{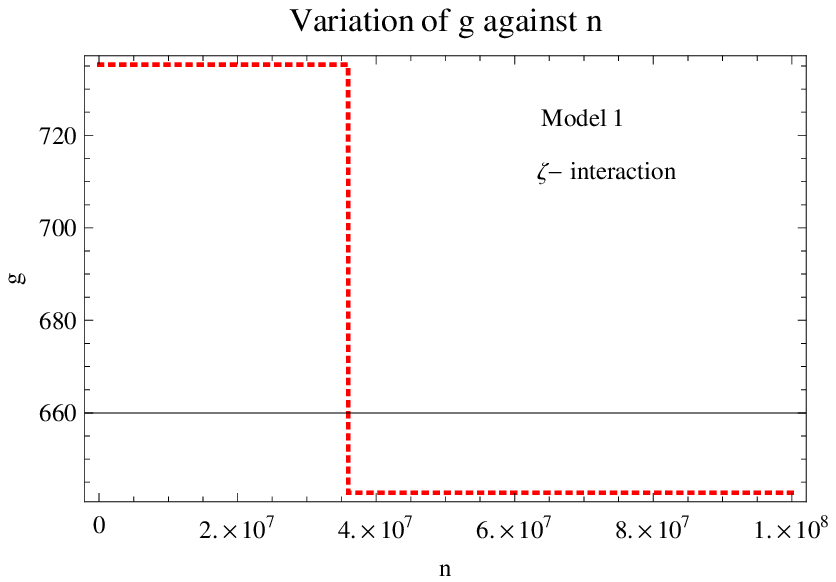}~~~\\
\vspace{5mm}
~~~~~~~~~~~~~~~~~~~~~~~~~~~Fig.1a~~~~~~~~~~~~~~~~~~~~~~~~~~~~~~~~~~~~~~~~~~~~~~~~~~~~~~~~~~~~Fig.1b~~~~~~~~~~~~\\
\vspace{5mm}
~~~~~~~~~~~~~~~~~~~~~~~~~~~~~~~~~~~~~~\includegraphics[height=2in]{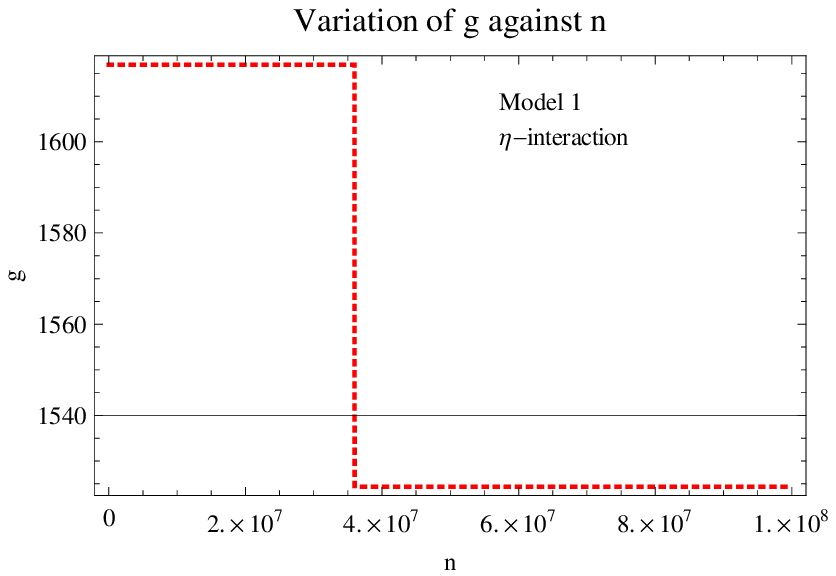}~~~~\\
\vspace{5mm}
~~~~~~~~~~~~~~~~~~~~~~~~~~~~~~~~~~~~~~~~~~~~~~~~~~~~~~~~~~~~~~Fig.1c~~~~~~~~~~~~~~~~~~~~~~~~~~~~~~~~~\\

\vspace{1mm} \textsl{Fig 1a : The plot of
$g(f_{0},H_{0},r_{s0},q_{0})$ against $n$ for model 1 using
$\Gamma$ interaction. The other parameters are considered as
$q=-0.2, r=3/7, H_{0}=72, \beta=0.1, \gamma=0.4,
\Gamma=0.5$}\\

\textsl{Fig 1b : The plot of $g(f_{0},H_{0},r_{s0},q_{0})$ against
$n$ for model 1 using $\zeta$ interaction. The other parameters
are considered as
$q=-0.2, r=3/7, H_{0}=72, \beta=0.1, \gamma=0.4, \zeta=0.5$}\\

\textsl{Fig 1c : The plot of $g(f_{0},H_{0},r_{s0},q_{0})$ against
$n$ for model 1 using $\eta$ interaction. The other parameters are
considered as $q=-0.2,
r=3/7, H_{0}=72, \beta=0.1, \gamma=0.4, \eta=0.5$}\\
\end{figure}

\subsubsection{Model 2}

Our second model is the model proposed in (Jamil, M. et. al.
2012). In this model, to avoid analytic and computation problems,
we choose a suitable expression for $f(T)$  which contains a
constant, linear and a non-linear form of torsion. The model is
given by
\begin{equation}\label{25}
f(T)=2C_1 \sqrt{-T} +\alpha T+C_2,
\end{equation}

where $\alpha$, $C_1$ and $C_2$ are arbitrary constants. It is
evident from the model that $C_{1}=0$ leads to Teleparallel
gravity. The combination of the first and the third term of the
model corresponds to the cosmological constant EoS in the
background of \textit{f(T)} gravity (Myrzakulov, R. 2011).
Likewise by shuffling terms, cosmologists have been able to set up
various models of \textit{f(T)} gravity possessing distinguishable
features. It is worth mentioning the model that we are dealing
with presently may have gathered its motivation from the model of
Veneziano ghost (Karami, K. et. al. 2013)

There are two basic advantages of this model as a result of which
it is preferentially chosen over other alternatives. The first one
is its simplicity with numerical computations and other one is the
results obtained from this model can be easily compared to the
corresponding results in general relativity. An analysis performed
by Capozziello et al in (Capozziello, S. et. al. 2011) showed that
by choosing $C_{1}=\sqrt{6}H_{0}\left(\Omega_{m0}-1\right)$,
$C_{2}=0$ and $\alpha=\Omega_{m0}$, it is possible to estimate the
parameters of the model as functions of Hubble parameter,
cosmographic parameters and matter density parameters. Here
$\Omega_{m0}=\frac{\rho_{m0}}{3H_{0}^{2}}$ represents the present
value of dimensionless matter density parameter. In accordance
with the current observational data $\Omega_{m0}=0.3$.

We fit this model in the constraint eqn.(\ref{21}) and plot the
function $g(f_{0},H_{0},r_{s0},q_{0})$ against the parameter
$\alpha$ in figs.(2a), (2b) and (2c) for $\Gamma$, $\zeta$ and
$\eta$ interaction terms respectively. From figure (2a) it is
evident that the constraint (\ref{21}) is satisfied for $\alpha
\approx -0.5$. Similarly in figs. (2b) and (2c), the constraint is
satisfied for $\alpha \approx -1$ and $\alpha \approx -0.6$
respectively. This implies that for these values of $\alpha$, the
model admits a late time stationary scenario as far as the ratio
of energy densities is concerned. Therefore in the background of
this model, and considering torsion contribution to be analogous
to dark energy a stationary scenario is successfully achieved for
various interaction terms. Hence this model is perfectly viable
according to the latest cosmological observations.

\vspace{2mm}
\begin{figure}
~~~~~~~~\includegraphics[height=2in]{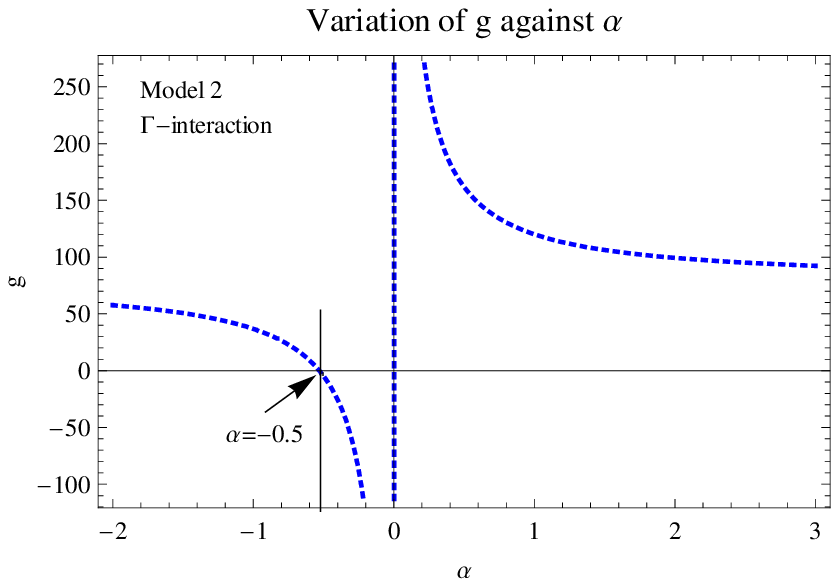}~~~~~~~~~~\includegraphics[height=2in]{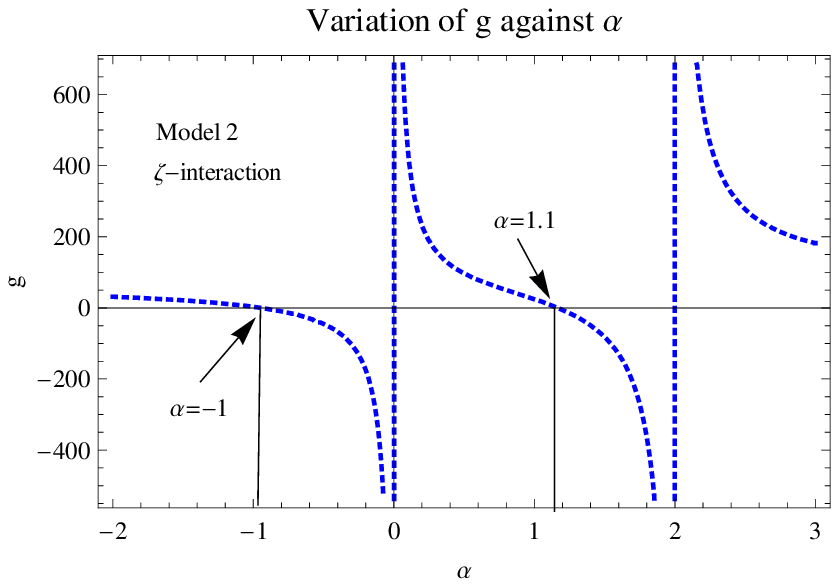}~~~~\\
\vspace{5mm}
~~~~~~~~~~~~~~~~~~~~~~~~~~~~~~~Fig.2a~~~~~~~~~~~~~~~~~~~~~~~~~~~~~~~~~~~~~~~~~~~~~~~~~~~~~~Fig.2b~~~~~~~~~~\\
\vspace{5mm}
~~~~~~~~~~~~~~~~~~~~~~~~~~~~~~~~~~~~~~~~~\includegraphics[height=2in]{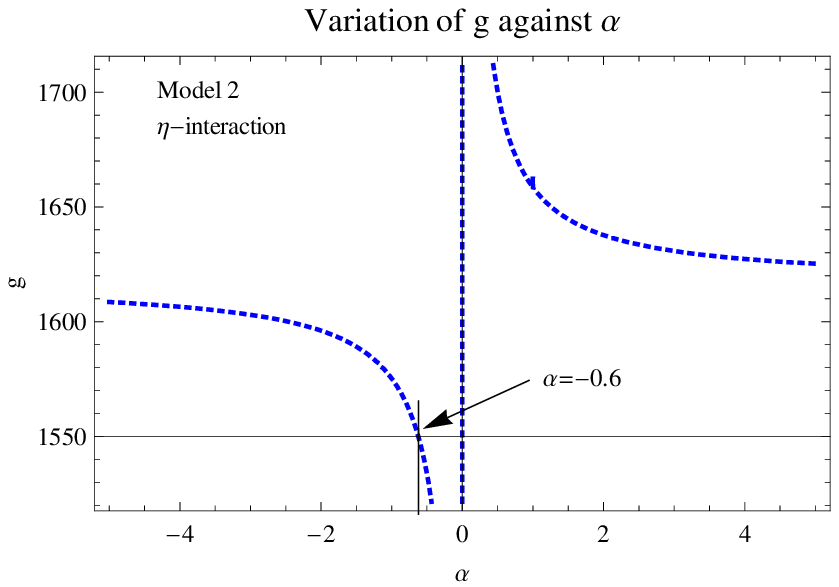}~~~~\\
\vspace{5mm}
~~~~~~~~~~~~~~~~~~~~~~~~~~~~~~~~~~~~~~~~~~~~~~~~~~~~~~~~~~~~~~~~~~~~Fig.2c~~~~~~~~~~~~~~~~~~~~~~~~~~~~~~~~~\\
\vspace{1mm} \textsl{Fig 2a : The plot of
$g(f_{0},H_{0},r_{s0},q_{0})$ against $\alpha$ for model 2 using
$\Gamma$ interaction. The other parameters are considered as
$q=-0.2, r=3/7, H_{0}=72, \Gamma=0.01,
C_{1}=\sqrt{6}H_{0}\left(\Omega_{m0}-1\right)=-123.4542830,
C_{2}=0$}\\

\textsl{Fig 2b : The plot of $g(f_{0},H_{0},r_{s0},q_{0})$ against
$\alpha$ for model 2 using $\zeta$ interaction. The other
parameters are considered as $q=-0.2, r=3/7, H_{0}=72,
C_{1}=\sqrt{6}H_{0}\left(\Omega_{m0}-1\right)=-123.4542830,
C_{2}=0, \zeta=0.5.$}\\

\textsl{Fig 2c : The plot of $g(f_{0},H_{0},r_{s0},q_{0})$ against
$\alpha$ for model 2 using $\eta$ interaction. The other
parameters are considered as $q=-0.2, r=3/7, H_{0}=72,
C_{1}=\sqrt{6}H_{0}\left(\Omega_{m0}-1\right)=-123.4542830,
C_{2}=0, \eta=0.5$}\\
\end{figure}

\vspace{5mm}

\subsubsection{Model 3 (Exponential Model)}

The model is given by (Linder, E. V. 2010; Bamba, K. et al 2011)
\begin{equation}\label{26}
f(T)=\alpha_{1}T\left(1-e^{p\frac{T_{0}}{T}}\right)
\end{equation}

where
$\alpha_{1}=-\frac{1-\Omega_{m}^{(0)}}{1-\left(1-2p\right)e^{p}}$
and $p$ is a constant. It must be noted that $p=0$ corresponds to
the $\Lambda CDM$ model. Moreover $T_{0}=T(z=0)$ is the torsion in
the present time. Here
$\Omega_{m}^{(0)}=\rho_{m}^{(0)}/\rho_{crit}^{(0)}$, where
$\rho_{m}^{(0)}$ is the energy density of non-relativistic matter
in the present time and $\rho_{crit}^{(0)}=\frac{3H_{0}^{2}}{8\pi
G}$ is the critical density. It is to be noted that the parameter
$\alpha_{1}$ that appears in the exponential model, has been
derived from
$\Omega_{DE}^{(0)}=\rho_{DE}^{(0)}/\rho_{crit}^{(0)}=1-\Omega_{m}^{(0)}=-\alpha\left[1-\left(1-2p\right)e^{p}\right]$
using the energy density of the effective dark energy of $f(T)$
theory. If $\Omega_{m}^{(0)}$ is given, then the exponential model
has a single parameter $p$. Using $p$ and $\Omega_{m}^{(0)}$ we
can calculate the values of other dimensionless quantities at
present time $(z=0)$. In (Bamba, K. et al 2011), we see that for
$p>0$, the universe enters the phantom regime, whereas for $p<0$,
the universe remains in the quintessence era.

We fit this model in the constraint eqn.(\ref{21}) and plot the
function $g(f_{0},H_{0},r_{s0},q_{0})$ against the parameter $p$
in figs.(3a), (3b) and (3c) for $\Gamma$, $\zeta$ and $\eta$
interaction terms respectively. From the figures it is seen that
for all the three interaction models the stationary scenario is
realized as far as the dark energy and dark matter is concerned.
For $\Gamma$, $\zeta$ and $\eta$ interaction models, stationary
scenario is realized for $p=0.15, p=0.30$ and $p=0.42$
respectively. From the figures it is clear that the coincidence
scenario is obtained only for positive values of $p$, i.e., when
the universe enters the phantom regime.

\vspace{5mm}
\begin{figure}
~~~~~~~~~~~~~~~\includegraphics[height=2in]{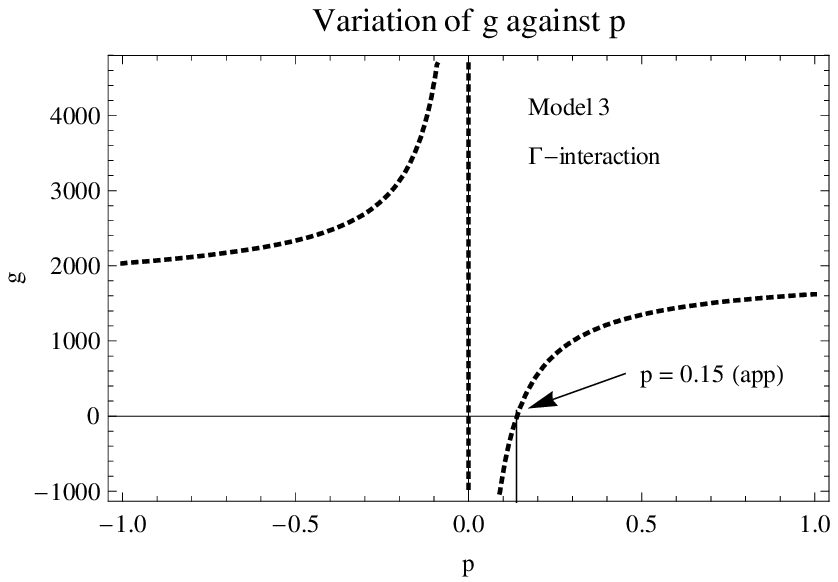}~~~~~~~~~~~~\includegraphics[height=2in]{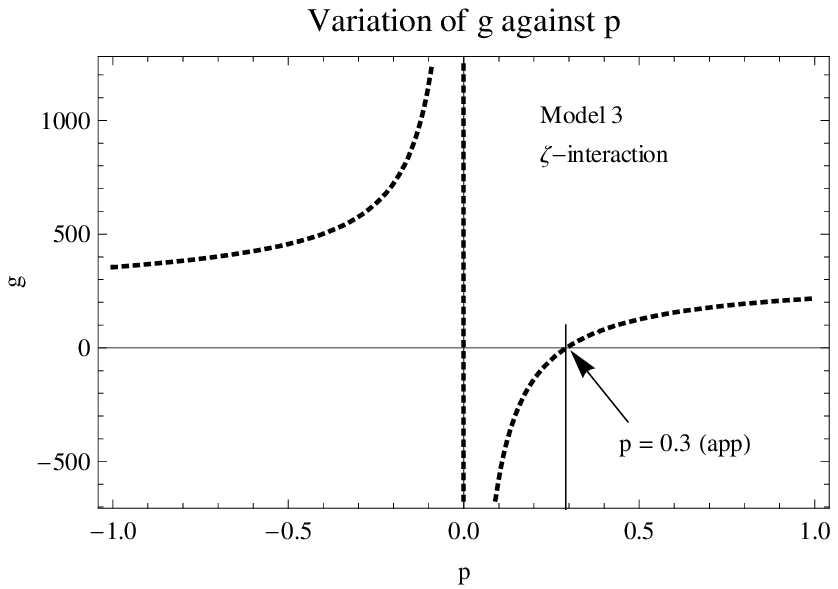}~~~~\\
\vspace{5mm}
~~~~~~~~~~~~~~~~~~~~~~~~~~~~~~~~~~~~~~Fig.3a~~~~~~~~~~~~~~~~~~~~~~~~~~~~~~~~~~~~~~~~~~~~~~~~~~~~Fig.3b~~~~~~~\\
\vspace{5mm}
~~~~~~~~~~~~~~~~~~~~~~~~~~~~~~~~~~~~~~\includegraphics[height=2in]{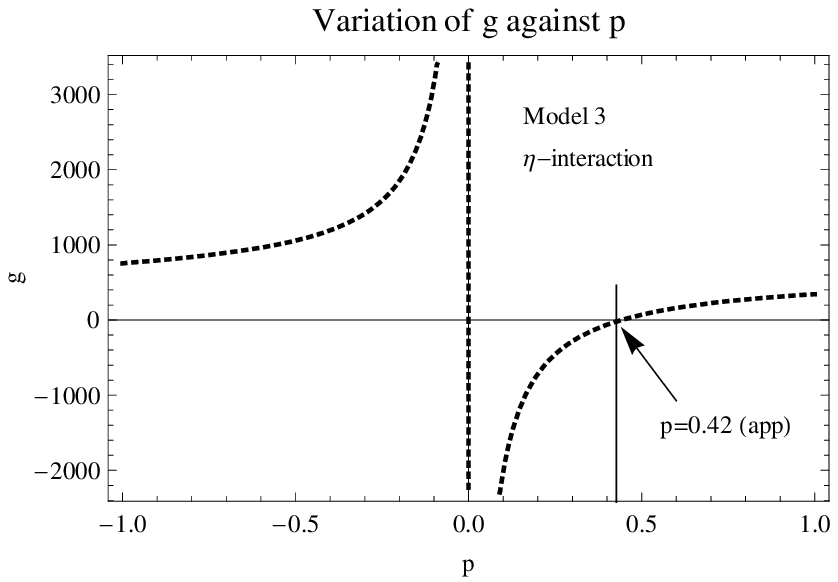}~~~~\\
\vspace{5mm}
~~~~~~~~~~~~~~~~~~~~~~~~~~~~~~~~~~~~~~~~~~~~~~~~~~~~~~~~~~~~~~~~~~~~~~~~Fig.3c~~~~~~~~~~~~~~~~~~~~~~~~~~~~~~~~~\\

\vspace{1mm} \textsl{Fig 3a : The plot of
$g(f_{0},H_{0},r_{s0},q_{0})$ against $p$ for model 3 using
$\Gamma$ interaction. The other parameters are considered as
$q=-0.2, r=3/7, H_{0}=72, \Gamma=5.$}\\

\textsl{Fig 3b : The plot of $g(f_{0},H_{0},r_{s0},q_{0})$ against
$p$ for model 3 using $\zeta$ interaction. The other parameters
are considered as
$q=-0.2, r=3/7, H_{0}=72, \zeta=3.$}\\

\textsl{Fig 3c : The plot of $g(f_{0},H_{0},r_{s0},q_{0})$ against
$p$ for model 3 using $\eta$ interaction. The other parameters are
considered as
$q=-0.2, r=3/7, H_{0}=72, \eta=3.$}\\
\end{figure}

\subsubsection{Model 4 (Logarithmic Model)}

The Logarithmic model of $f(T)$ theory is given by (Bamba, K. et
al 2011)
\begin{equation}
f(T)=\beta_{1}T_{0}\left(\frac{\epsilon
T_{0}}{T}\right)^{-1/2}\ln\left(\frac{\epsilon T_{0}}{T}\right)
\end{equation}
where $\beta_{1}\equiv \frac{1-\Omega_{m}^{(0)}}{2q^{-1/2}}$ and
$\epsilon$ is a positive constant. It must be noted that if
$\Omega_{m}^{(0)}$ is obtained in the same way as in the
exponential $f(T)$ model, then $\epsilon >0$ is the only parameter
of the logarithmic model. It was found in (Bamba, K. et al 2011)
that equation of state $(\omega)$ never crosses the phantom line
for this model.

On fitting this model in the constraint eqn.(\ref{21}) we get the
corresponding plot for the function $g(f_{0},H_{0},r_{s0},q_{0})$
against the parameter $\epsilon$ in figs.(4a), (4b) and (4c) for
$\Gamma$, $\zeta$ and $\eta$ interaction terms respectively. From
the figures it is seen that for all the three interaction models
the coincidence scenario is realized between dark energy and dark
matter. For $\Gamma$, $\zeta$ and $\eta$ interaction models,
stationary scenario is realized for $\epsilon=3950, \epsilon=4670$
and $\epsilon=2800$ respectively. It must be noted here that
stationary scenario in this case is realized only for negative
value of interaction coupling constant. This shows that the
direction of flow is just the reverse compared to the previous
models, i.e., from dark energy to dark matter.

\vspace{2mm}
\begin{figure}
~~~~~~~~\includegraphics[height=2in]{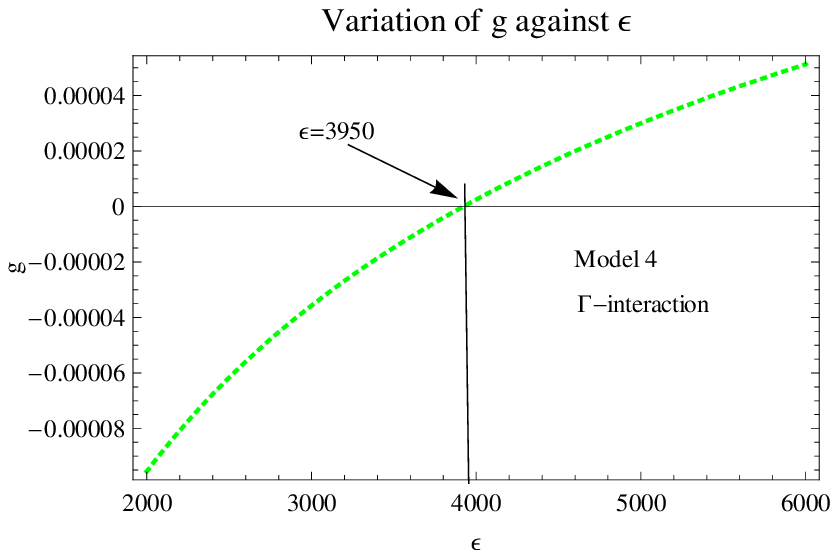}~~~~~~~~~~~~\includegraphics[height=2in]{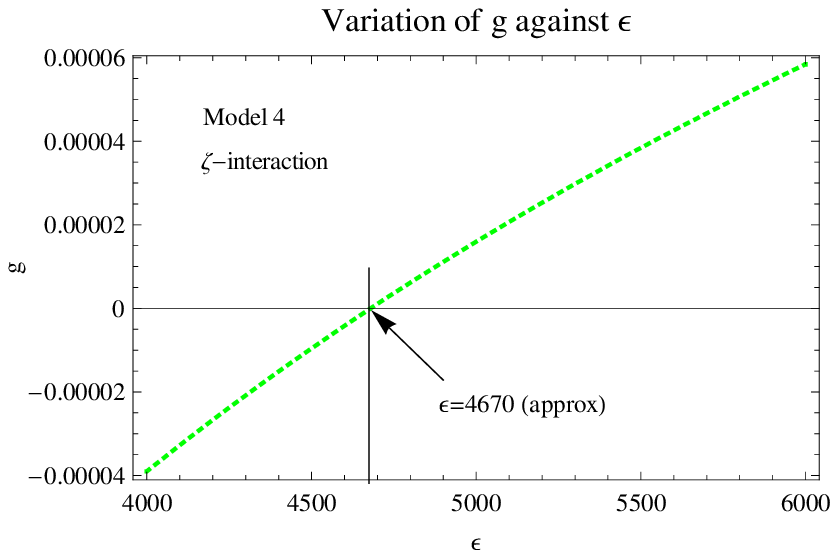}~~~~\\
\vspace{5mm}
~~~~~~~~~~~~~~~~~~~~~~~~~~~~~~~Fig.4a~~~~~~~~~~~~~~~~~~~~~~~~~~~~~~~~~~~~~~~~~~~~~~~~~~~~~~Fig.4b~~~~~~~~~~~~~~~~~~~\\
\vspace{5mm}
~~~~~~~~~~~~~~~~~~~~~~~~~~~~~~~~~~~~~~\includegraphics[height=2in]{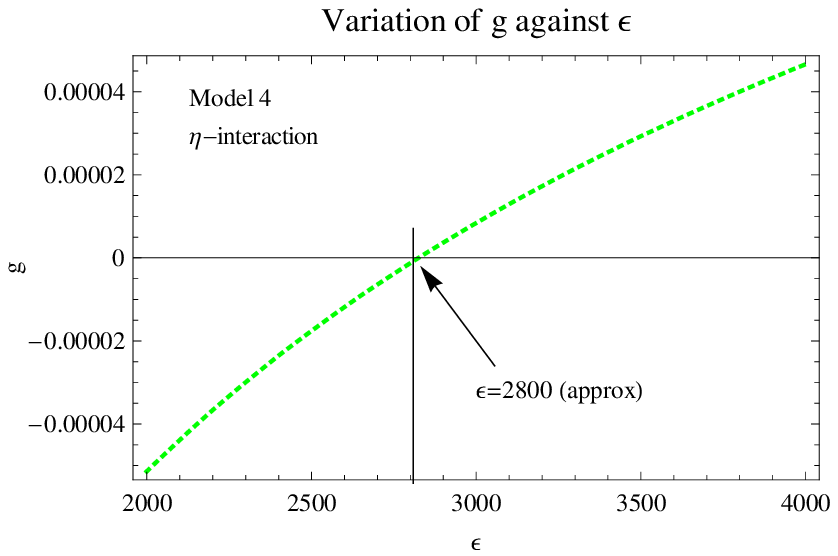}~~~~\\
\vspace{5mm}
~~~~~~~~~~~~~~~~~~~~~~~~~~~~~~~~~~~~~~~~~~~~~~~~~~~~~~~~~~~~Fig.4c~~~~~~~~~~~~~~~~~~~~~~~~~~~~~~~~~\\

\vspace{1mm} \textsl{Fig 4a : The plot of
$g(f_{0},H_{0},r_{s0},q_{0})$ against $T/T_{0}$ for model 4 using
$\Gamma$ interaction. The other parameters are considered as
$q=-0.2, r=3/7, H_{0}=72, \Gamma=-.167998$}\\

\textsl{Fig 4b : The plot of $g(f_{0},H_{0},r_{s0},q_{0})$ against
$T/T_{0}$ for model 4 using $\zeta$ interaction. The other
parameters are considered as
$q=-0.2, r=3/7, H_{0}=72, \zeta=-.559987.$}\\

\textsl{Fig 4c : The plot of $g(f_{0},H_{0},r_{s0},q_{0})$ against
$T/T_{0}$ for model 4 using $\eta$ interaction. The other
parameters are considered as
$q=-0.2, r=3/7, H_{0}=72, \eta=-.239997.$}\\
\end{figure}

\subsubsection{Model 5}
The model is given by (Myrzakulov, R. 2011)
\begin{equation}
f(T)=\alpha_{2}T+\frac{\beta_{2}}{T}
\end{equation}
where $\alpha_{2}$ and $\beta_{2}$ are constants.

Here also we see that the constraint eqn.(\ref{21}) is satisfied
and a satisfactory stationary scenario is achieved between dark
energy and dark matter. From the figs. (5a), (5b) and (5c), we see
that the coincidence is attained for all the three interaction
models $\Gamma$, $\zeta$ and $\eta$ at $\alpha_{2}=950,
\alpha_{2}=1015$ and $\alpha_{2}=1350$ respectively. Just like the
previous model, we see that in this model also a negative value of
coupling constant is necessary to attain the stationary scenario.
Therefore, here also the direction of flow is from dark energy to
dark matter.

\vspace{2mm}
\begin{figure}
~~~~~~~~\includegraphics[height=2in]{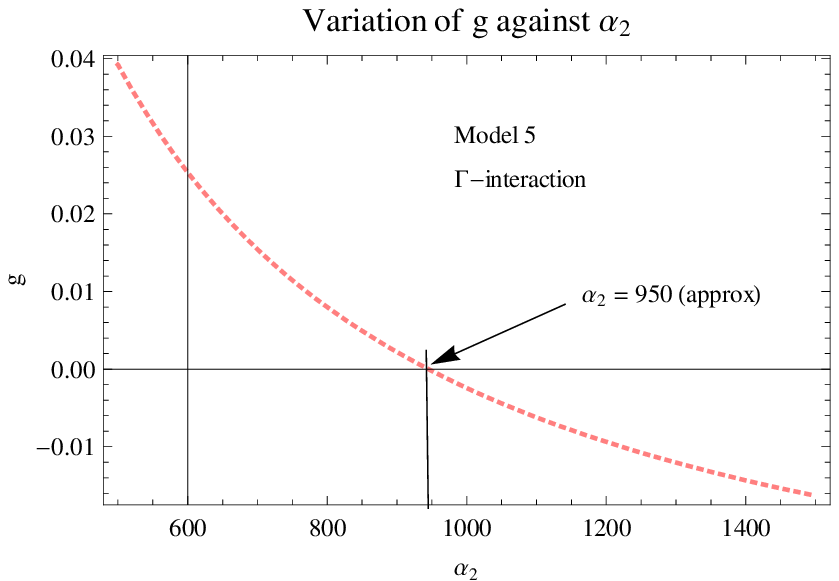}~~~~~~~~~~~~\includegraphics[height=2in]{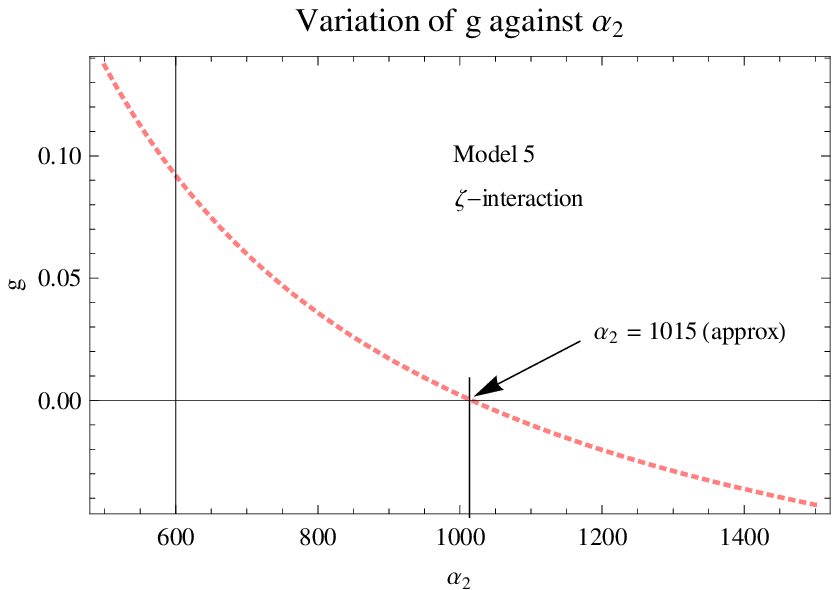}~~~~\\
\vspace{5mm}
~~~~~~~~~~~~~~~~~~~~~~~~~~~~~~~Fig.5a~~~~~~~~~~~~~~~~~~~~~~~~~~~~~~~~~~~~~~~~~~~~~~~~5b~~~~~~~~~~~\\
\vspace{5mm}
~~~~~~~~~~~~~~~~~~~~~~~~~~~~~~~~~~~~~~\includegraphics[height=2in]{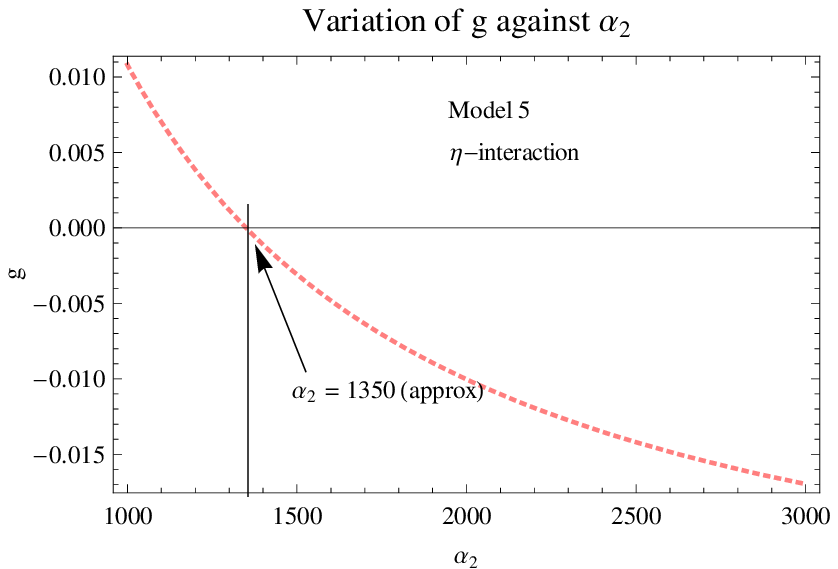}~~~~\\
\vspace{5mm}
~~~~~~~~~~~~~~~~~~~~~~~~~~~~~~~~~~~~~~~~~~~~~~~~~~~~~~~~~~~~Fig.5c~~~~~~~~~~~~~~~~~~~~~~~~~~~~~~~~~\\
\vspace{1mm} \textsl{Fig 5a : The plot of
$g(f_{0},H_{0},r_{s0},q_{0})$ against $\alpha_{2}$ for model 5
using $\Gamma$ interaction. The other parameters are considered as
$q=-0.2, r=3/7, H_{0}=72, \beta=2, \Gamma=-0.1681.$}\\

\textsl{Fig 5b : The plot of $g(f_{0},H_{0},r_{s0},q_{0})$ against
$\alpha_{2}$ for model 5 using $\zeta$ interaction. The other
parameters are considered as
$q=-0.2, r=3/7, H_{0}=72, \beta=2, \zeta=-0.561$}\\

\textsl{Fig 5c : The plot of $g(f_{0},H_{0},r_{s0},q_{0})$ against
$\alpha_{2}$ for model 5 using $\eta$ interaction. The other
parameters are considered as
$q=-0.2, r=3/7, H_{0}=72, \beta=2, \eta=-0.2401.$}\\

\end{figure}

\section{Conclusion}
In this assignment we have devised a method to probe the
contribution of torsion, interacting with dark matter to realize
the present scenario of cosmic coincidence. The comparable values
of the densities of dark energy (energy from torsion component)
and dark matter have been attributed to the presence of a suitable
interaction between them. The choice of the interaction term
obviously was not unanimous. So we decided to consider three
different forms of possible interaction terms widely found in
literature. A setup was designed and a constraint equation was
formed which will filter the models that accommodate a stationary
scenario in its framework, from those which do not. This will help
us to differentiate the cosmologically viable models from the
others. To demonstrate the designed set-up, we considered five
specific models of \textit{f(T)} gravity available in literature,
using all the three forms of interactions. It was found that the
first model showed a considerable departure from the stationary
scenario. On the contrary the other four models were perfectly
consistent with our setup and generated a satisfactory stationary
scenario, thus showing their cosmological viability and their
superiority over their counterparts. For the third model
(exponential model) it was seen that the cosmological coincidence
is realized only in the phantom regime, which is a very
interesting result. For the fourth (logarithmic model) and the
fifth models, we see that the the stationary scenario is attained
for negative interaction values. This shows that the direction of
flow must be from dark energy to dark matter unlike the previous
models. Under such circumstances the universe will return from the
present energy dominated phase to a matter dominated phase. It can
also be speculated that this phenomenon may analogically
correspond to a future deceleration (Chakraborty et. al. 2014; Pan
et. al. 2015) or even a cosmic contraction. This result is quite
unexpected. The reason behind this is not quite clear and we keep
it an open question for the time being. Moreover, it must be
mentioned that the realization of the stationary scenario for the
fourth and the fifth models was quite restrained. From the
numerical simulations it was found that the scenario could only be
achieved for the given values of the interaction coupling constants.
So its clear that the coincidence phenomenon actually constrains
the interaction parameters for these models.\\\\

\section*{Acknowledgements}
The Author acknowledges the anonymous referee
for enlightening comments that helped to improve the quality of the manuscript.\\

\section*{Compliance with Ethical Standards}
The author declares that he has no conflict of interest.

\end{document}